\documentclass[letterpaper, 10 pt, conference]{ieeeconf}
\IEEEoverridecommandlockouts    
\usepackage{graphics} 
\usepackage{epsfig} 
\usepackage{times} 
\usepackage{amsmath} 
\usepackage{amssymb}  
\usepackage{subfigure}
\usepackage{mathtools}
\usepackage{color}
\usepackage{cite}
\usepackage{comment}
\usepackage{balance}
\usepackage[table]{xcolor}
\usepackage{multirow}
\usepackage{hyperref}
\hypersetup{
    colorlinks=true,
    linkcolor=black,
    citecolor=green,
    urlcolor=blue,
}
\usepackage{blindtext}
\usepackage{bm}
\usepackage{algorithm}
\usepackage{algpseudocode}
\usepackage[official]{eurosym}

\DeclareMathOperator{\diag}{diag}

\algrenewcommand\algorithmicindent{1.0em}
\algnewcommand\algorithmicswitch{\textbf{switch}}
\algnewcommand\algorithmiccase{\textbf{case}}
\algnewcommand\algorithmicassert{\texttt{assert}}
\algnewcommand\Assert[1]{\State \algorithmicassert(#1)}%
\algdef{SE}[SWITCH]{Switch}{EndSwitch}[1]{\algorithmicswitch\ #1\ \algorithmicdo}{\algorithmicend\ \algorithmicswitch}%
\algdef{SE}[CASE]{Case}{EndCase}[1]{\algorithmiccase\ #1}{\algorithmicend\ \algorithmiccase}%
\algtext*{EndSwitch}%
\algtext*{EndCase}%

\makeatletter
\newcommand{\algmargin}{\the\ALG@thistlm}
\makeatother
\newlength{\forwidth}
\algdef{SE}[parFOR]{parFor}{EndparFor}[1]
  {\parbox[t]{\dimexpr\linewidth-\algmargin}{%
     \hangindent\forwidth\strut\algorithmicfor\ #1\ \algorithmicdo\strut}}{\algorithmicend\ \algorithmicfor}%
\newlength{\forif}
\algnewcommand{\parState}[1]{\State%
  \parbox[t]{\dimexpr\linewidth-\algmargin}{\strut #1\strut}}

\newtheorem{theorem}{Theorem}[section]
\newtheorem{remark}{Remark}[section]

\newtheorem{assumption}{Assumption}[section]

\newtheorem{problem}{Problem}[section]

\newenvironment{hidden}{}{}

\title{\LARGE \bf
Beyond Line-of-Sight: Cooperative Localization \\ Using Vision and V2X Communication
}

\author{Annika Wong, Zhiqi Tang, Frank J. Jiang, Karl H. Johansson, Jonas Mårtensson
\thanks{
    $^{1}$Division of Decision and Control Systems, EECS, KTH Royal Institute of Technology, Malvinas v{\"a}g 10, 10044 Stockholm, Sweden {\tt\small \{ytawong, ztang2, frankji, kallej, jonas1\}@kth.se}.}%
}

\begin{document}

\maketitle
\thispagestyle{empty}
\pagestyle{empty}

\begin{abstract}
Accurate and robust localization is critical for the safe operation of Connected and Automated Vehicles (CAVs), especially in complex urban environments where Global Navigation Satellite System (GNSS) signals are unreliable. This paper presents a novel vision-based cooperative localization algorithm that leverages onboard cameras and Vehicle-to-Everything (V2X) communication to enable CAVs to estimate their poses, even in occlusion-heavy scenarios such as busy intersections. 
In particular, we propose a novel decentralized observer for a group of connected agents that includes landmark agents (static or moving) in the environment with known positions and vehicle agents that need to estimate their poses (both positions and orientations). Assuming that (i) there are at least three landmark agents in the environment, (ii) each vehicle agent can measure its own angular and translational velocities as well as relative bearings to at least three neighboring landmarks or vehicles, and (iii) neighboring vehicles can communicate their pose estimates, each vehicle can estimate its own pose using the proposed decentralized observer. We prove that the origin of the estimation error is locally exponentially stable under the proposed observer, provided that the minimal observability conditions are satisfied.
Moreover, we evaluate the proposed approach through experiments with real 1/10th-scale connected vehicles and large-scale simulations, demonstrating its scalability and validating the theoretical guarantees in practical scenarios.

\end{abstract}

\section{Introduction}\label{sec:intro}

\begin{figure}
     \centering
     \subfigure[]{
    \includegraphics[width=0.45\linewidth]{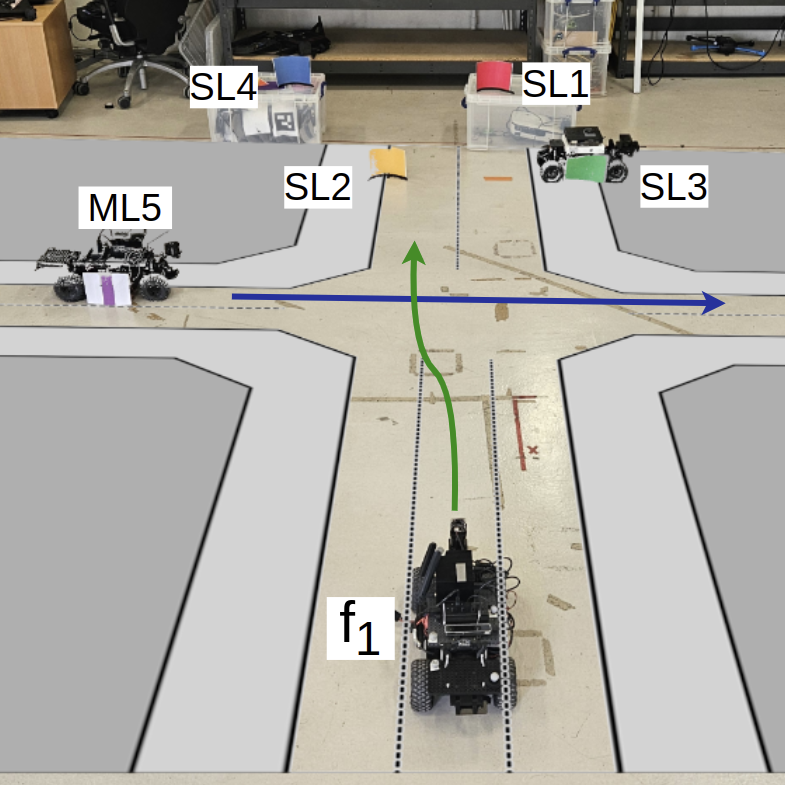}
    \label{fig:crossing_path_intersection_setup}
     }
     \hfill
     \subfigure[]{
    \includegraphics[width=0.45\linewidth]{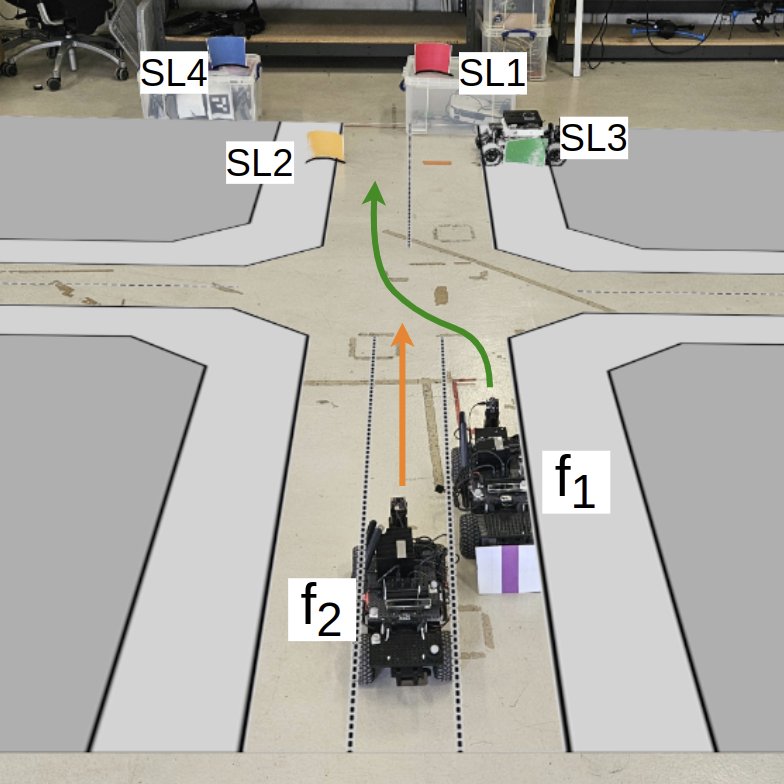}
    \label{fig:overtaking_intersection_setup}
     }
    \vspace{-0.5em}
    \caption{Experimental setup for (a) crossing path and (b) overtaking scenarios. SL and ML refer to static and moving landmarks, respectively, belonging to the landmark agents. $f_1$ and $f_2$ denote the vehicles estimating their poses, belonging to the vehicle agents. 
    The arrows indicate the intended trajectories of the agents.
    }
    \label{fig:experimental_setup}
\end{figure}

Precise localization is fundamental for the safe operation of Connected and Automated Vehicles (CAV). The need for precise localization is even more dire in urban intersections due to high traffic density, dynamic interactions between diverse road users, and frequent changes in the surrounding environment. For example, in Fig. \ref{fig:experimental_setup}, we illustrate two simple intersection scenarios where two vehicles need to safely maneuver together in close quarters. These conditions demand highly precise and consistent localization, as even minor pose estimation errors can lead to unsafe maneuvers or collisions. 

Over the last decade, many have proposed and developed a variety of localization approaches for the operation of CAVs. One of the common approaches to obtain positioning in outdoor scenarios is to utilize Global Navigation Satellite Systems (GNSS) due to their availability throughout the transportation network \cite{GNSSAndVisualOdometry2016}. However, GNSS is not always consistently precise or reliable enough for CAV operation, particularly in urban areas. Signal accuracy is often degraded due to issues such as multi-path interference caused by the surrounding infrastructure (i.e. urban canyons) and reception loss in GNSS-denied environments, such as tunnels and underground passages \cite{IntroToGNSS2021}.
Due to the limitations of 
GNSS, including its unreliability in certain conditions and its inability to provide information about the surrounding environment, it is often necessary to 
employ alternative sensors. These typically include dedicated onboard exteroceptive sensors, such as cameras or lidar, in combination with proprioceptive sensors like IMUs or wheel encoders. 
By doing so, we can develop localization methods for CAVs that perform precisely and reliably enough for safe, autonomous driving.   


Recently, vision-based localization of a single ego vehicle has seen growing attention, and one of the conventional approaches is to match detected visual features such as lane markings, traffic lights, and identifiable landmarks with a pre-defined map \cite{VisionCompareLaneAndCameraGPS2018, FrontFacingCameraAndLaneCompareWithPF2016, MonocularCameraMapCompare2018, IntersectionWithTrafficLightMap2019}. 
However, critical challenges still persist such as low robustness, limited detection precision, and difficulties in deployment in complex urban environment where the occlusion of landmark features often occur in busy traffic scenarios such as interactions, as illustrated in Fig.~\ref{fig:experimental_setup}. Motivated by the rapid development of 5G/6G technology for vehicle-to-vehicle (V2V) and vehicle-to-infrastructure (V2I) communication, there are several proposals to leverage cooperative localization to overcome these challenges \cite{FuzzyLogic2014, BayesianApproachCooperative2015, EKFandCI2016,RangeonlyWithSICF2018, FactorGraph2016}. By utilizing information provided by nearby vehicles and infrastructure, each ego-vehicle could enhance its own localization accuracy and robustness beyond the capabilities of individual onboard sensors. However, the mentioned approaches often suffer from practical challenges stemming from the scaling issues of centralized fusion, over-reliance on communication reliability, and lack of formal stability guarantees on the localization error.  


In this paper, we address these challenges by taking inspiration from the recent success of bearing-based visual localization methods~\cite{zhao2016bearing,Selazo2014}. In these methods, vehicles equipped with onboard cameras can measure their neighboring vehicles and landmarks' relative bearings (directions) as long as they are in the cameras' field of view. Under the assumption that neighboring vehicles can also communicate their position estimations, the positions of a group of vehicles can be estimated. Due to their robustness and computational efficiency, bearing-based approaches hold strong potential for addressing many of the practical challenges currently encountered in vision-based cooperative localization. However, existing works for generalized bearing-based cooperative localization~\cite{tang2022localization, tang2024observer} require additional theoretical development and practical considerations before they can be deployed on CAVs, which we address in this work. Specifically, our contributions are as follows:
\begin{enumerate}
    \item we extend the concept of \textit{Bearing Persistently Exciting} sensor networks \cite{tang2021formation} to cover practical situations where bearings can only be measured in each vehicle's camera frame, 
    \item we propose a novel decentralized observer for estimating the poses (positions and orientations) of large-scale CAVs, where each vehicle can measure its linear and angular velocities, obtain visual bearings to neighboring vehicles and landmarks, and receive pose estimates of its neighbors through communication, 
    \item we prove that the origin of the estimation error is exponentially stable under the proposed observer, provided that the minimal observability conditions are satisfied, 
    \item we validate and showcase the robustness and performance of the proposed cooperative localization approach, even under occlusion conditions, on both real hardware and in larger-scale simulations.
\end{enumerate}
Through this development, we combine the benefits of vision and communication for a cooperative localization approach that is both precise and robust. Furthermore, the implementation of our proposed approach is publicly available.\footnote{\href{https://github.com/Annika-wyt/Cooperative-Localization}{https://github.com/Annika-wyt/Cooperative-Localization}}

The remainder of the paper is organized as follows. In Section~\ref{sec:prelim}, we provide some preliminaries for our approach. In Section~\ref{sec:mot_ex}, we describe the problem formulation considered throughout the paper. In Section~\ref{sec:method}, we present the vision-based cooperative localization framework. In Section~\ref{sec:exp}, we validate the performance of our approach via simulations and experiments. In Section~\ref{sec:conc}, we conclude the paper with a discussion about our work and future directions.

\section{Preliminaries}\label{sec:prelim}

In this section, we introduce some preliminaries that we need before explicitly formulating our problem and presenting our cooperative localization approach. We start by overviewing the basic ideas of bearing-based localization and the key concepts that we will build upon in this paper. Then, we will detail some notations and graph theory used in the remainder of the paper.

\subsection{Bearing-Based Cooperative Localization}
 Given the limited communication bandwidth and limited cameras' field of view, the key challenge of bearing-based cooperative localization is ensuring connected agents' observability under reduced sensing and intermittent communication topologies.
Researchers have previously investigated bearing rigidity theory, which defines the minimum interaction topology required to determine the observability of sensor networks using constant inter-agent bearing measurements \cite{zhao2016bearing,Selazo2014}. The work in \cite{tang2021formation,tang2022relaxed} has recently generalized the results by proposing the concepts of \textit{Bearing Persistently Exciting} (BPE) sensor networks and \textit{Relaxed Bearing Rigidity} (RBR), which leverages time-varying bearings to relax the minimum graph topologies required in classical bearing rigidity. It is promising to exploit the concepts of BPE and RBR for cooperative localization of CAVs, particularly in dynamic traffic scenarios. However, existing work only considered simplified scenarios where all the vehicles are able to measure relative bearings in a common inertial frame without considering orientation estimation \cite{tang2022localization, tang2024observer}. This is not applicable in practice, where vehicles measure bearings from their own camera frame and both position and orientation need to be estimated. We address this practical limitation in this paper, but before that, we introduce the necessary notation and graph theory for developing our solution.

\begin{figure}
    \centering
    \vspace{0.5em}
    \includegraphics[width=0.7\linewidth]{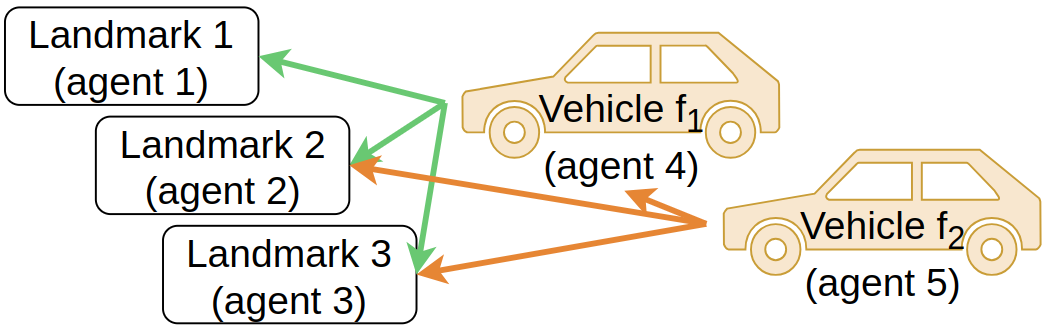}
    \caption{Possible interaction topology under Assumption \ref{assum:graph}.}
    \label{fig:possible_graph}
    \vspace{-1em}
\end{figure}

\subsection{Notation}
We denote by $\mathbb S^2:=\{y\in\mathbb{R}^3:\|y\|=1\}$ the $2$-Sphere and $\|.\|$ the Euclidean norm. 
  For any $y\in \mathbb S^2$, we define the projection operator $\Pi_y := I - y y^{\top} \geq 0$
which projects any vector $y\in\mathbb{R}^3$ to the plane orthogonal to $y$.
$S(\cdot)$ represents the skew-symmetric matrix such that $S(x)y = x \times y$, $\forall x\in \mathbb R^3, y \in \mathbb R^3$. 
Denote $\diag(A_i) = blkdiag(A_1,\ldots, A_k)\in \mathbb R^{3k\times 3k}$ the block diagonal matrix with elements given by $A_i\in \mathbb R^{3\times 3}$.

\subsection{Graph theory}
We model the underlying interaction topology between CAVs and visual landmarks (both static and moving) as a digraph (directed graph) $\mathcal{G} := (\mathcal{V}, \mathcal{E})$, where $\mathcal{V}=\{1,2,\ldots,n\}$ is the set of vertices and $\mathcal{E} \subseteq \mathcal{V} \times \mathcal{V}$ is the set of directed edges. The set of neighbors of agent $i$ is denoted by $\mathcal{N}_i:=\{j\in\mathcal{V}|(i,j)\in\mathcal{E}\}$. Define $m_i=|\mathcal{N}_i|$, where $|.|$ denotes the cardinality of a set.

In this work, we consider agents under a directed leader-follower structure under the following assumption:
\begin{assumption}\label{assum:graph}
$\mathcal{G}=(\mathcal V,\mathcal E)$ is a directed acyclic graph and has at least three root vertices. $\mathcal V$ contains two types of vertices, i.e.,  $\mathcal V=\mathcal V_l\cup \mathcal V_f$, where $\mathcal V_l$ represents landmark vertices and  $\mathcal V_f$ denotes vehicle vertices. Without loss of generality, we assume that $\mathcal V_l=\{1,\ldots,n_l\}$ and $\mathcal V_f=\{n_l+1,\ldots,n_l+n_f\}$ with $n=n_l+n_f$. The landmark agents are root vertices with no neighbors, i.e., $\mathcal N_i=\emptyset, i\in \mathcal V_l$. For vehicle agents $i\in \mathcal V_f$, $\mathcal N_i\subseteq \{1, \ldots, i-1\}$.
\end{assumption}

In Fig. \ref{fig:possible_graph}, we illustrate an example of a possible interaction topology between two CAVs and three landmarks. The arrow from agent $i$ to $j$ ($j\in \mathcal N_i$) indicates that agent $i$ can sense relative information with respect to $j$ and receive information that agent $j$ communicates.


\section{Problem Formulation}\label{sec:mot_ex}



In this paper, we address the cooperative localization problem for a group of connected and automated vehicles operating in congested traffic scenarios—for example, at intersections where accurate localization is crucial, but where occlusions may degrade the localization performance of an individual ego vehicle.
We do this by extending the concept of BPE sensor networks under a directed leader-follower interaction topology \cite{tang2021formation} by considering practical scenarios that relative bearing measurements are measured in each vehicle's camera frame. 


Formally speaking, in a congested traffic scenario, we assume there are $n=n_l+n_f$ agents, including $n_l$ landmark agents and $n_f$ vehicle agents. We assume that the $n_f$ vehicle agents do not know their pose information and are equipped with an onboard camera, wheel encoders, and an IMU.  The landmark agents can be static, such as feature points on nearby infrastructure; they can also be nonstatic, for instance, moving vehicles with known inertial position. 

Let $p_i \in \mathbb{R}^3$ denote the position of agent $i\in \mathcal V$ expressed in its body frame $\mathcal{B}_i$ with dynamics 
\begin{equation}\label{eq:dot_bar_p}
    \dot{p}_i = -S(\omega_i)p_i + v_i
\end{equation}
where $\omega_i \in \mathbb{R}^3$ is the angular velocity of $\mathcal{B}_i$ expressed in $\mathcal{B}_i$ and $v_i \in \mathbb{R}^3$ is the translational velocity of agent $i$ expressed in $\mathcal{B}_i$. The body frame $\mathcal{B}_i$ of the agent $i$ with respect to the common inertial frame $\mathcal{F}$ is represented by the orientation matrix $R_i \in SO(3)$ with the following kinematic relations:
\begin{equation}\label{eq:dot_R}
    \dot R_i = R_iS(\omega_i)
\end{equation}
The bearing vector of agent $j$ with respect to agent $i$ expressed in $\mathcal{B}_i$ is defined as: 
\begin{equation}
 g_{ij} := \frac{p_i-R_i^\mathsf{T} \bar p_j}{||p_i-R_i^\mathsf{T} \bar p_j||} \in \mathbb S^2  
\end{equation}
where  $\bar {p}_j:=R_jp_j$ denote the position of agent $j$, $j \in \mathcal{N}_i$, expressed in $\mathcal F$. If agent $i$ is equipped with an onboard camera and agent $j$ is in its camera's field of view, bearing $g_{ij}$ can be obtained using the image coordinate $ p'_{ij}=[p_x \ p_y \ 1]^\top$
\begin{equation}
    \label{eq:bearing_calculation}
    g_{ij} = \frac{p'_{ij}}{||p'_{ij}||}
\end{equation}

In this work, we make the following assumptions about the visual landmark agents in the environment:
\begin{assumption}\label{ass:three_landmarks}
    There are at least three (static or moving) landmark agents (i.e., $n_l\ge 3$) in the $n$-agent sensor network with known position vector $\bar p_i, i \in \mathcal{V}_l$. 
\end{assumption}

Additionally, we also assume the localization approach is developed in an Intelligent Transport Systems (ITS) context where there are roadside units that facilitate V2I and V2V communication. In this context, we impose the following practical assumptions on the sensing and communication between agents.
\begin{assumption} \label{ass:construction}
    The sensing and communication topology of the $n$-agent sensor network is described by a graph topology $\mathcal G$ defined in Assumption \ref{assum:graph}. Each vehicle agent $i$'s $ (i\in \mathcal V_f)$ onboard camera can measure the relative bearing vectors $g_{ij}$ to its neighbors $ j\in \mathcal{N}_i$, the onboard wheel encoders and IMU can measure its translational velocity vector $v_i$ and angular velocity $\omega_i$, and the onboard V2X communication device can get receive estimates $(\hat R_j, \hat p_j)$ of its neighboring agents.
\end{assumption}

We also assume boundness and well-definess of the state variables as follows:
\begin{assumption}\label{ass:desired} 
    The velocities of each agent $v_{i}(t) \text{ and } \omega_i(t),\ \forall i\in \mathcal V$ are bounded and the inter-agent bearings $g_{ij}(t),\ \forall (i,j)\in \mathcal E$ are well-defined $\forall t\ge0$, i.e., $p_i(t)$ and $p_{j}(t)$ are not collocated, $\forall t>0, \forall i,j\in \mathcal V, i\ne j$.
\end{assumption}

With these theoretical and practical assumptions, we formulate the problem as follows.
\begin{problem}\label{prob:main}
    Under the Assumptions \ref{ass:three_landmarks} - \ref{ass:desired}, design real-time decentralized localization algorithms using relative bearings $g_{ij}$, angular velocity $\omega_i$ and translational velocity $v_i$ measurements to estimate the pose $(R_i, p_i)$ of each vehicle agent $i \in \mathcal{V}_f$.
\end{problem}

\section{Vision-Based Cooperative Localization}\label{sec:method}

\begin{figure}
    \centering
    \vspace{0.5em}
    \includegraphics[width=\linewidth]{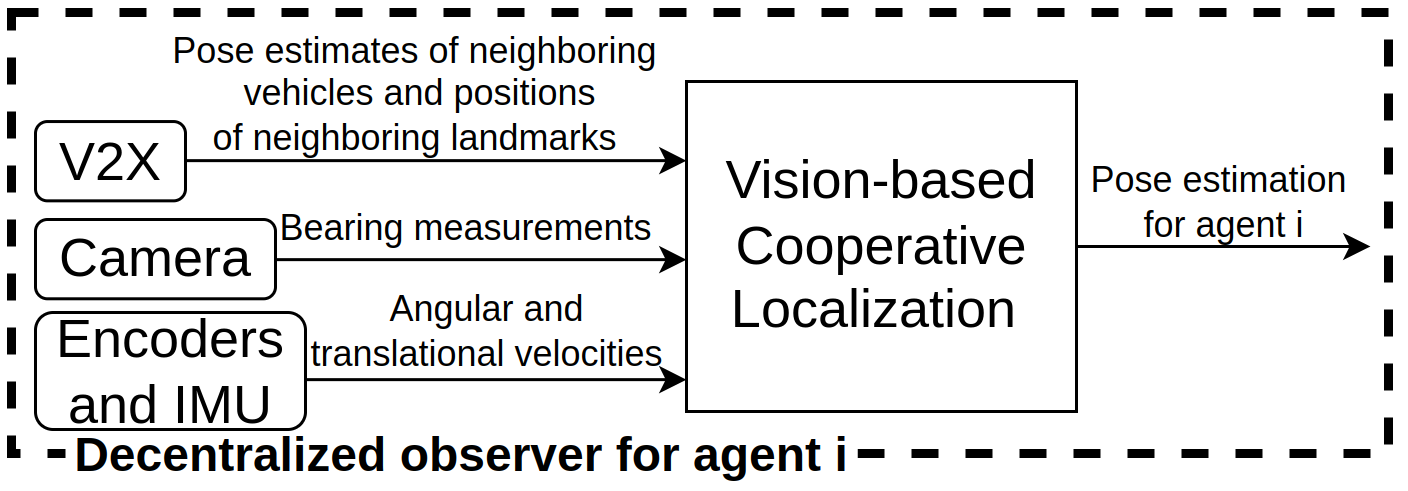}
    \caption{Overview of the vision-based cooperative localization for agent $i \in \mathcal{V}_f$.}
    \label{fig:system-arch}
    \vspace{-1em}
\end{figure}

\begin{hidden}
\end{hidden}


In this section, we present our approach to solving Problem~\ref{prob:main}. Based on the bearing-based Riccati observer design using nonstationary Perspective-n-Point for a single agent \cite{hamel2017riccati}, we propose a novel decentralized observer for a group of connected agents which contains landmark agents (static or moving) with known positions and vehicle agents that need to estimate their poses. 
We illustrate an overview of the design of each CAV's observer in Fig.~\ref{fig:system-arch}.

To start, let $(\hat R_i, \hat{p}_i)$ denote an estimate of the pose $(R_i, p_i)$. For a landmark agent $i$, $i \in \mathcal{V}_l$, we assume that $\hat{p}_i = p_i$ and $\hat R_i = R_i$. For a vehicle agent $i, i \in \mathcal{V}_f$, define the dynamics of $(\hat R_i, \hat{p}_i)$ as,
    \begin{equation}\label{dot_estmate}
    \begin{aligned}
    \dot {\hat R}_i&=\hat R_iS(\hat \omega_i)\\
    \dot {\hat {p}}_i&=-S(\omega_i)\hat{p}_i + \hat {v}_i
    \end{aligned}
    \end{equation}
where $\hat \omega_i$ and $\hat {v}_i$ are the estimated angular and translational velocities to be designed later. 

Define the orientation errors as $\tilde{R}_i:=\hat{R}_i^\mathsf{T}R_i$.
The unit quaternion associated to $\tilde{R}_i$ is defined as $\Lambda_i=(\mathring{\lambda}_i,\lambda_i)$ with $\lambda_i \in \mathbb{R}^3$ (resp. $\mathring{\lambda}_i\in \mathbb R^1$) denoting the vector (resp. scalar) part of the quaternion. Using the corresponding Rodrigues formula relating $\Lambda_i$ to $\tilde{R}_i$ and the property of unit quaternion as in Appendix \ref{appex:rotation_body}, one can parameterize $\tilde R_i$ as a function of $\lambda_i$
\begin{equation}
\tilde{R}_i = I_3+2S(\lambda_i)+O(|\lambda_i|^2)
\end{equation}
where $O(\cdot)$ is the higher order term. The dynamics of $\lambda_i$ satisfies
\begin{equation}\label{eq:Inertialdotlambdabody}
\begin{aligned}
2\dot{\lambda}_i = - S(\omega_i)(2\lambda_i) +\tilde{\omega}_i  +O(|\lambda_i||\tilde \omega_i|) +O(|\lambda_i|^2)
\end{aligned}
\end{equation}
 Under this parameterization, we can model the dynamics of the pose $(2\lambda_i, p_i)$ as 
\begin{equation} \label{eq:dynamics-body}
\begin{aligned}
 \begin{bmatrix}
    2\dot{\lambda}_i\\
    \dot{p}_i 
 \end{bmatrix}=A_i(t)\begin{bmatrix}
    2{\lambda}_i\\
    {p}_i 
 \end{bmatrix} +\begin{bmatrix}\tilde{\omega}_i +O(|\lambda_i||\tilde \omega_i|) +O(|\lambda_i|^2)\\v_i\end{bmatrix}
\end{aligned} 
\end{equation}
where
\begin{equation}
A_i(t) = \begin{bmatrix}
        -S(\omega_i) & 0_{3} \\
        0_{3} & -S(\omega_i) 
\end{bmatrix}
\end{equation}

Since each vehicle agent $i\in \mathcal V_f$ can measure the relative bearings with respect to its neighbors $j \in \mathcal{N}_i$ and receive neighbor $j$'s estimated position, the measurement output function of vehicle $i$ with respect to neighbor $j$ is designed as 
\begin{equation}
y_{ij}=\Pi_{g_{ij}}\hat R_i^\mathsf{T} \hat R_j \hat p_j
\end{equation}

Define $y_i: = [y_{ij}^\mathsf{T} \cdots y_{ik}^\mathsf{T}]^\mathsf{T} \in \mathbb{R}^{3m_i}, \forall j,k \in \mathcal{N}_i, j \neq k$ as the stacked measurement of vehicle $i$ including all relative bearings with respect to its neighboring agents. Using the representation of $y_{ij}$ in the Appendix \ref{appex:y-body}, we can represent $y_i$ as following function

\begin{equation}
\begin{aligned}
    y_{i} =& C_{i,1}2\lambda_i + C_{i,2}p_i - D_{i} \\&+ O(|p_j||\lambda_i|^2) + O(|\bar p_j-\hat R_j\hat p_j||\lambda_i|), \forall j \in \mathcal{N}_i
    \end{aligned}
\end{equation}
where
\begin{equation*}
    C_{i,1} = \begin{bmatrix}
        -\Pi_{g_{ij}}S(\hat R_i^\mathsf{T} \hat R_j \hat p_j) \\
        \vdots \\
        -\Pi_{g_{ik}}S(\hat R_i^\mathsf{T} \hat R_k \hat p_k) 
    \end{bmatrix} \in \mathbb{R}^{3m_i \times 3}
\end{equation*}
\begin{equation*}
    C_{i,2} = \begin{bmatrix}
        \Pi_{g_{ij}} \\
        \vdots \\
        \Pi_{g_{ik}}
    \end{bmatrix} \in \mathbb{R}^{3m_i \times 3} 
\end{equation*}
\begin{equation}\label{eq:D_body}
    D_i = \begin{bmatrix}
        \Pi_{g_{ij}}\hat R_i^\mathsf{T} (\bar p_j-\hat R_j\hat p_j) \\
        \vdots \\
        \Pi_{g_{ik}}\hat R_i^\mathsf{T}(\bar p_k-\hat R_k\hat p_k) \\
    \end{bmatrix} \in \mathbb{R}^{3m_i \times 3}
\end{equation}
$\forall j, k \in \mathcal{N}_i, j \neq k$ and recalling $m_i = |\mathcal{N}_i|$.

\vspace{0.5em}
Define matrix $K_i = k_iP_iC_i^\mathsf{T}Q_i = \begin{bmatrix}
    K_{i,1} \\
    K_{i,2}
\end{bmatrix}\in \mathbb R^{6\times 3m_i}$ to be used later in the observer design. $k_i > 0.5$ is a bounded gain, $P_i$ is the solution of the following Continuous Riccati Equation: $\dot P_i = A_iP_i + P_iA_i^\mathsf{T} - P_iC_i^\mathsf{T} Q_iC_iP_i + V_i$ with $P_i(0) \in \mathbb{R}^{6\times6}$ a symmetric positive definite matrix, $Q_i = \diag\{q_{ij}I_3\} \in \mathbb{R}^{3m_i\times 3m_i}$, $j \in \mathcal{N}_i$ with $q_{ij} > 0$ positive weights, $V_i \in \mathbb{R}^{6\times6}$, a bounded continuous symmetric positive definite matrix-valued function, and $C_i(\lambda_i, \hat p_i, t) = \begin{bmatrix}
    C_{i,1} & C_{i,2}
\end{bmatrix}$. Design the estimated angular and translational velocities as 

\begin{equation}\label{eq:observer_body}
    \left\{\begin{aligned}
        \hat \omega_i = \omega_i + K_{i,1}(y_i - C_{i,2}\hat {p}_i) \\
        \hat{v}_i = v_i + K_{i,2}(y_i - C_{i,2}\hat {p}_i) \\
    \end{aligned}\right.
\end{equation}
where $(\hat \omega_i,\hat v_i)$ is used to compute the estimated pose $(\hat R_i,\hat p_i)$ from \eqref{dot_estmate} via numerical integration.
\begin{theorem}\label{theorem:mobilevelocity}
    Consider an $n$-agent ($n\ge 4$) system $\mathcal{G}(\boldsymbol{p}(t))$ in $\mathbb{R}^3$ space under dynamics \eqref{eq:dot_bar_p} and \eqref{eq:dot_R}. Assume that Assumptions \ref{ass:three_landmarks} - \ref{ass:desired} are satisfied. Under the decentralized observer \eqref{eq:observer_body}, the origin of the estimation error $(\lambda_i, \tilde {p}_i)$ is locally exponentially stable (ES) for all vehicle agents $\forall i \in \mathcal{V}_f$, provided $Q_i$ is larger than some positive matrix and $(A_i, C^*_i)$ is uniformly observable with
    \begin{equation*}
        C^*_i = C_i(0, p_i, t) = \begin{bmatrix}
        -\Pi_{g_{ij}}S(R_i^\mathsf{T} \bar p_j) & \Pi_{g_{ij}}  \\
        \vdots & \vdots \\
        -\Pi_{g_{ik}}S(R_i^\mathsf{T} \bar p_k) & \Pi_{g_{ik}} 
        \end{bmatrix}
    \end{equation*}
    $\forall j, k \in \mathcal{N}_i \text{ and, } j \neq k$.
\end{theorem}

The proof of the Theorem is in Appendix \ref{appex:proof}.
\begin{remark}\label{remark}
    Following a similar discussion in \cite{hamel2017riccati}, the uniform observability condition is guaranteed for most of the situations when each vehicle agent $i, i\in \mathcal V_f$ has at least three neighbors whose positions are not aligned. There are some special cases that violate the conditions. A formal analysis of the observability condition is beyond the scope of this paper and is left for future work.
\end{remark}


\section{Experimental Results}\label{sec:exp}

In this section, we validate the theoretical guarantees of the proposed cooperative localization approach with real hardware in the loop. To easily evaluate a variety of scenarios, we choose to evaluate our approach on a 1/10th-scale intersection using the Small Vehicles for Autonomy (SVEA) platform \cite{jiang2022svea} at the Integrated Transport Research Lab at the KTH Royal Institute of Technology. We start by introducing the experimental setup and then present the results of our experimental evaluation in three scenarios, showcasing the benefits of the developed cooperative localization approach. For more specific details about the experimental setup, we refer readers to the source code in the footnote of page 2.




\subsection{Experimental setup}

We set up a 1/10th-scale four-way intersection including landmarks, SVEA vehicles, and an edge server.
Color markers are placed in the environment and serve as static landmarks. In practice, they could be replaced by traffic signs or other features on the nearby infrastructure. 
A color marker is also placed on a SVEA for vehicle detection. 
The SVEAs are equipped with an onboard NVIDIA Jetson Orin computer and a Pixy2 camera. 
To facilitate V2X communication, we connect the SVEAs using an edge server,
over indoor WiFi for ease of scenario development.


 To simplify the procedure for landmark and vehicle detection, we employ a color-based filtering method. 
Upon detection of the color marker, the Pixy2 camera provides the image coordinates and ID of the detected marker at 60 Hz. The ID is used to associate the bearing measurement with the known position of the corresponding landmark. Using the image coordinates and the intrinsic camera matrix, the bearing of the detected object is then computed using \eqref{eq:bearing_calculation}. 
The translational and angular velocities of each vehicle agent are estimated based on the control inputs to the actuators at 100 Hz. Given that the vehicles operate on a 2D surface, it is assumed that the angular velocities about the x- and y-axes remain zero throughout the experiments. In practical scenarios, translational and angular velocity measurements could be obtained from wheel encoders and IMUs. 


All the vehicle agents $i$, $\forall i \in \mathcal{V}_f$, run the proposed observer \eqref{eq:observer_body} onboard using the Robot Operating System (ROS) at 60 Hz, to estimate the pose of its body frame $\mathcal{B}_i$, which is transformed from the camera frame by a translational factor. 
Qualisys motion capture system (MOCAP) is used to compare the estimation with the ground truth. We will focus on the estimated position errors in the common inertial frame $\tilde {\bar p}_i:=\bar p_i - \hat R_i\hat p_i$ instead of in the body frame in the following plots. 



\begin{hidden}

\end{hidden}

\subsection{Experimental Scenarios and Results}
To evaluate the theoretical convergence guarantees and resiliency of the approach to occlusions, we consider three different scenarios.
For all vehicle agents $i$, $\forall i \in \mathcal{V}_f$, the initial pose estimation $(\hat R_i(0),\hat p_i(0))$ is random and inaccurate, which reflects the potential inaccuracy of GNSS in an urban canyon. For clarity of index notation, we denote $f_k = n_l + k, k\ge 1$ as the vertex index of the $k$th vehicle agent for the following discussion. 
\subsubsection{\textbf{Crossing Path Scenario}} 
as illustrated in Fig.~\ref{fig:crossing_path_intersection_setup}, the vehicle that creates an occlusion travels eastbound while vehicle $f_1$ moves northbound through the intersection. This configuration represents a typical intersection case where vehicles approach from different directions, often resulting in brief occlusions due to overlapping paths. 
In this scenario, we consider a $6$-agent system: there are four static landmarks and one moving landmark (i.e., the vehicle drives eastbound) with known position, and one vehicle agent $f_1$ needs to estimate its pose, as shown in Fig.~\ref{fig:crossing_path_intersection_setup}. The gains are chosen as $k_{f_1}=80$, $q_{f_1 j}=25\quad \forall j \in \mathcal{N}_{f_1}$, $V_{f_1}=\diag\{0.1I_3, I_3\}$, and $P_{f_1}(0)=\diag\{I_3, 10I_3\}$.

\begin{figure}
     \centering
    \vspace{0.5em}
      \subfigure[]{
         \includegraphics[width=0.6\linewidth]{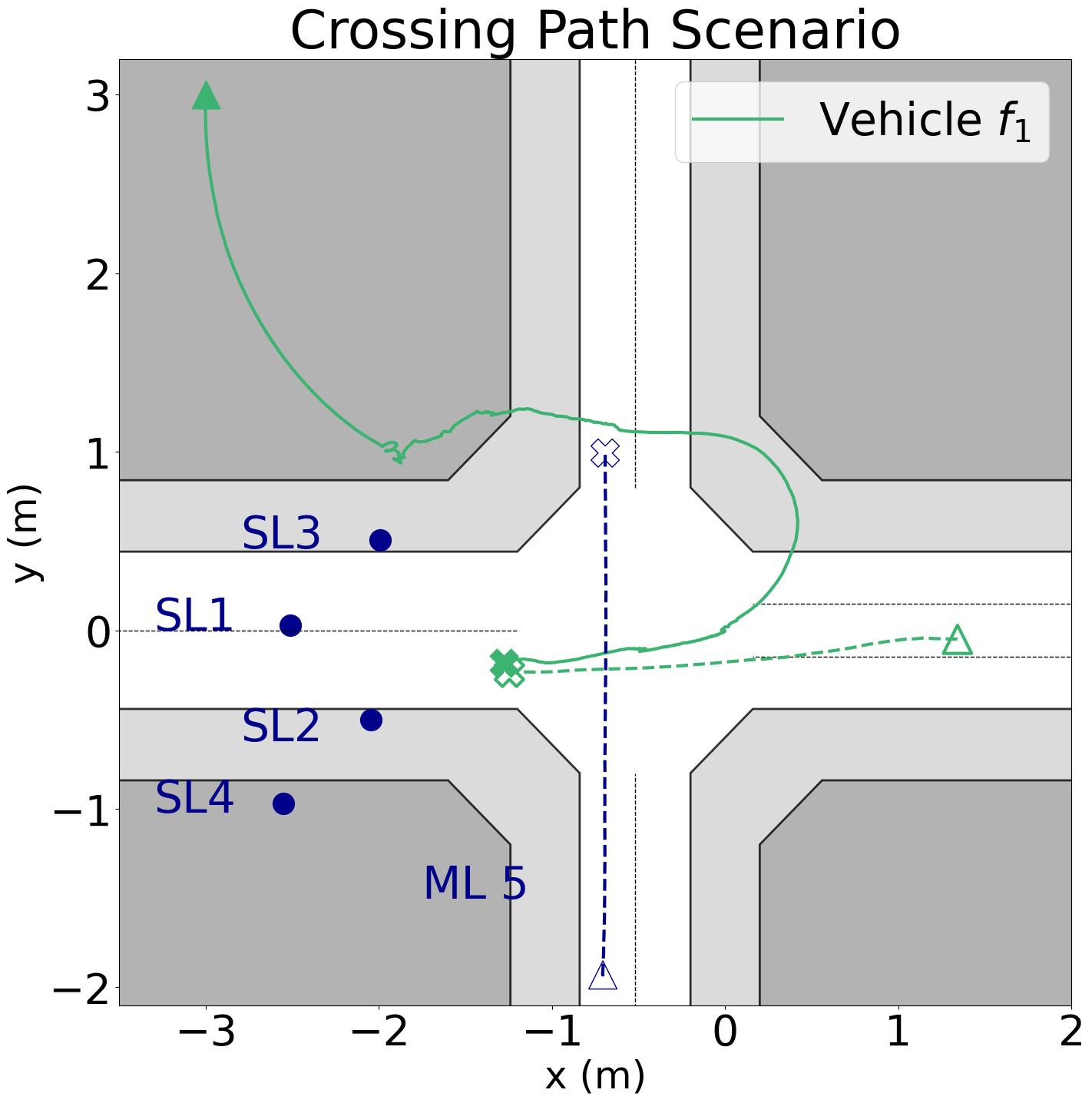}
        \label{fig:body_crossing_path_2d}
         }
     \begin{minipage}[t]{1\columnwidth}
     \subfigure[]{
        \includegraphics[width=0.45\linewidth]{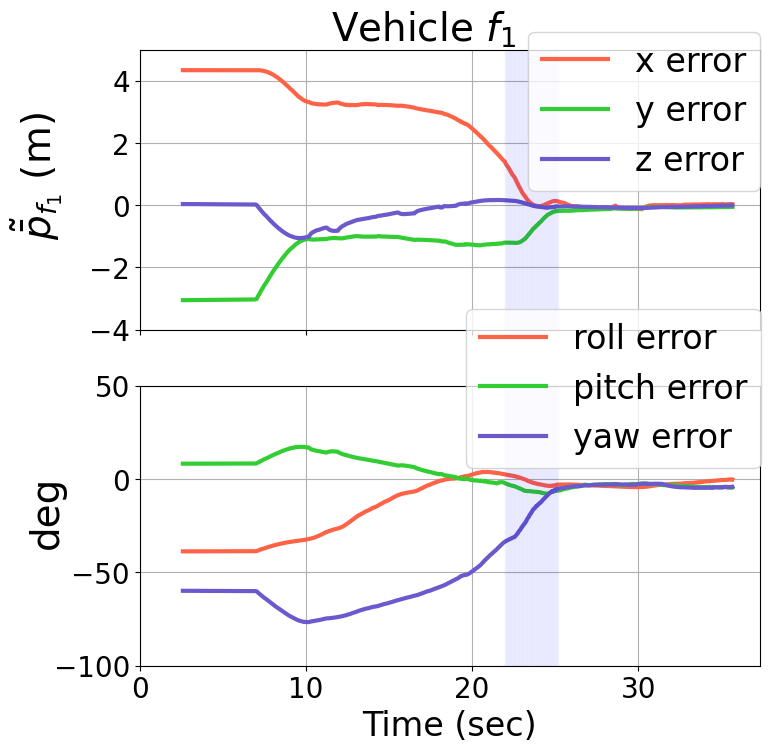}
        \label{fig:body_crossing_path_f1}
        }
        \subfigure[]{
            \includegraphics[width=0.45\linewidth]{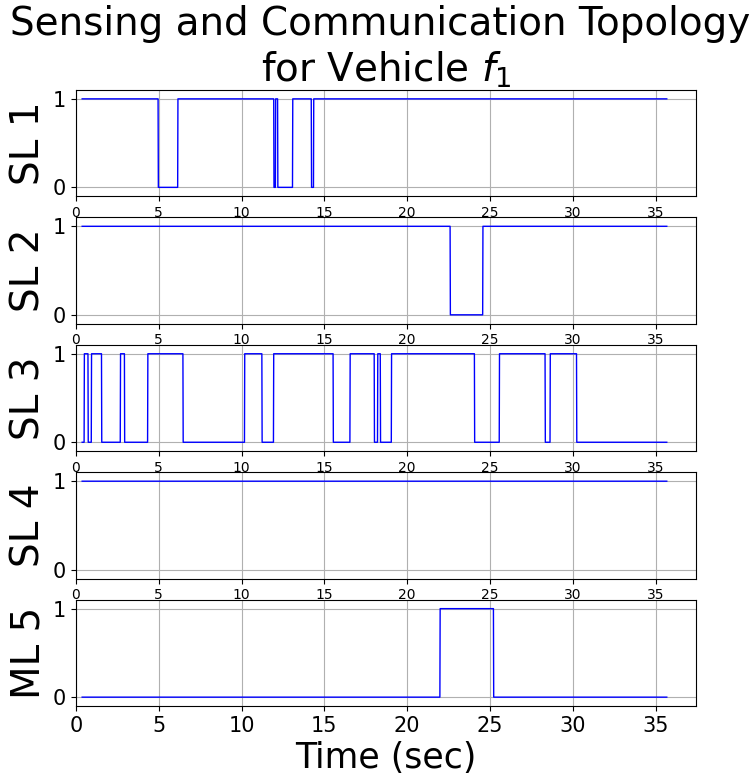}
            \label{fig:body_crossing_landmark_detection}
        }
     \end{minipage}
    \vspace{-1em}
    \caption{(a) The projected 2D trajectories of the vehicle $f_1$ and the moving landmark in the crossing path scenario, with triangles and crosses indicating the initial and the final positions, respectively, dotted and solid lines indicating the ground truth and estimation, respectively, SL and ML refer to the static and moving landmarks, respectively.
    (b) Evolution of error states (position in common inertial frame $\mathcal{F}$ and orientation) of vehicle $f_1$. (c) Sensing and communication topology of vehicle $f_1$.
    }
    \label{fig:body_crossing_path}
\end{figure}

Fig. \ref{fig:body_crossing_path_2d} shows the projected 2D trajectories of the vehicle $f_1$'s position estimation and ground truth. Note that although the ground truth position is in the 2D space, the initial estimate is chosen freely in 3D space. Fig. \ref{fig:body_crossing_path_f1} shows the evolution of estimation errors throughout the experiments, with the blue-shaded areas representing when the moving landmark was detected. We conclude that the state estimation errors converge fast enough to zero under large initial errors, with the presence of disturbance and noise from sensors. 
Fig. \ref{fig:body_crossing_landmark_detection} indicates that the interaction topology of vehicle $f_1$ is switching graph topologies, due to the limited camera's field of view and occlusion.
When the moving landmark was detected from 21s to 25s by vehicle $f_1$, occlusion happened, with the detection of SL2 and SL4 intermittently lost, see Fig. \ref{fig:body_crossing_landmark_detection}. However, the detection of the moving landmark helped to maintain the vehicle's observability, ensuring that vehicle $f_1$ could continue estimating its pose without degradation in accuracy. Note that we have verified in the experiment that our observers work well under switching graph topology, as long as the observability condition (Remark \ref{remark}) is satisfied at each time instant.
The video of the experimental result can be found at: \href{https://bit.ly/crossing-path-scenario}{https://bit.ly/crossing-path-scenario}.


\begin{figure}
     \centering
    \vspace{0.5em}
     \subfigure[]{
        \includegraphics[width=0.6\linewidth]{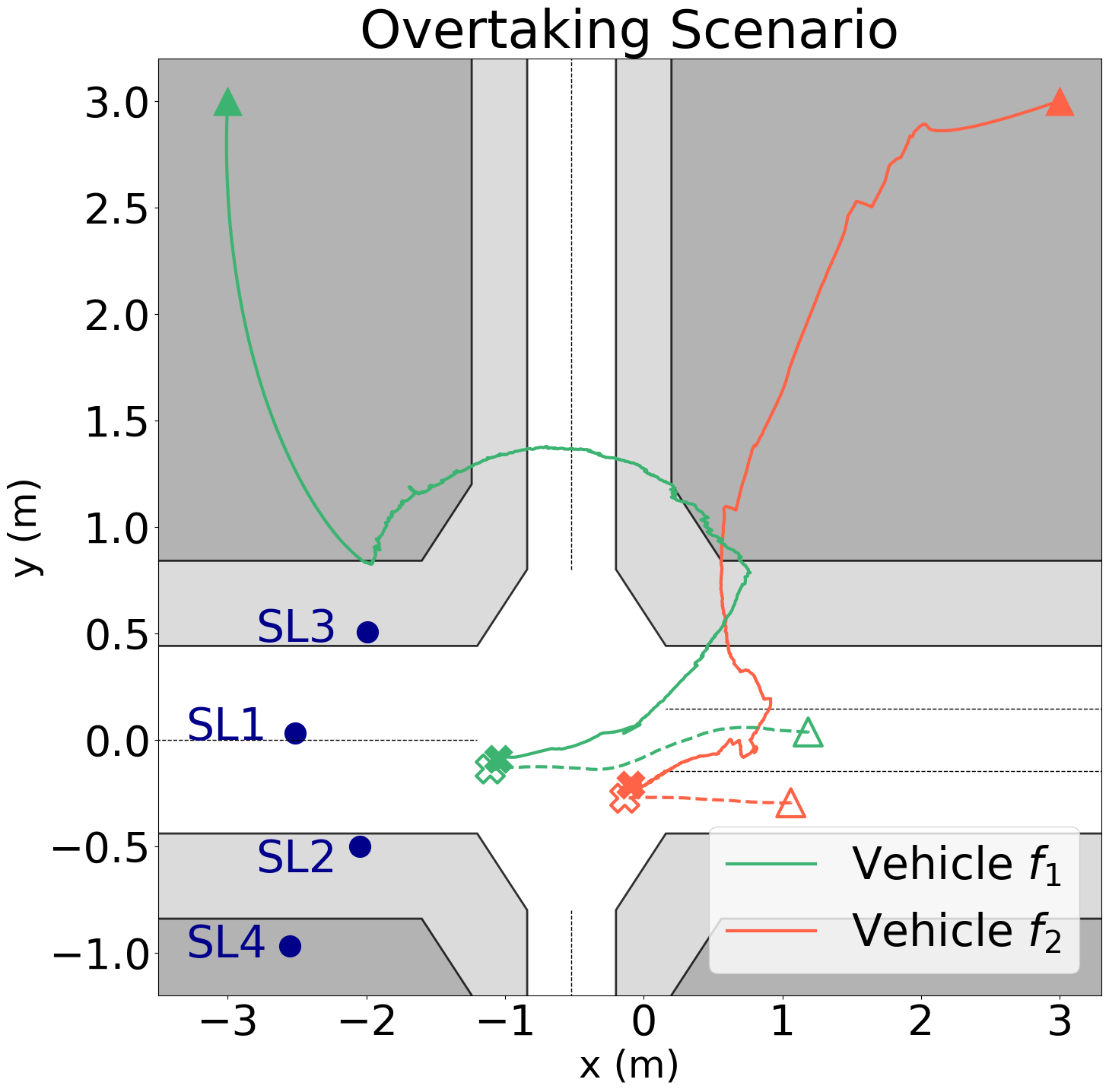}
        \label{fig:body_overtake_2d}
     }
     \begin{minipage}[t]{1\columnwidth}
        \subfigure[]{
            \includegraphics[width=.45\linewidth]{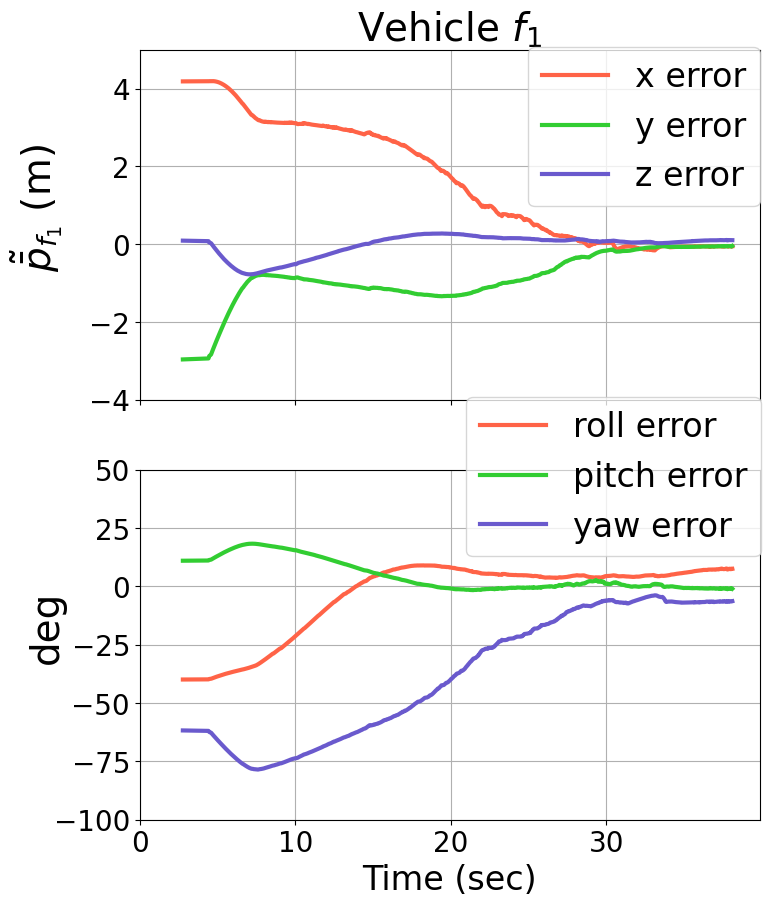}
            \label{fig:body_overtake_f1}
        }
         \subfigure[]{
         \includegraphics[width=0.45\linewidth]{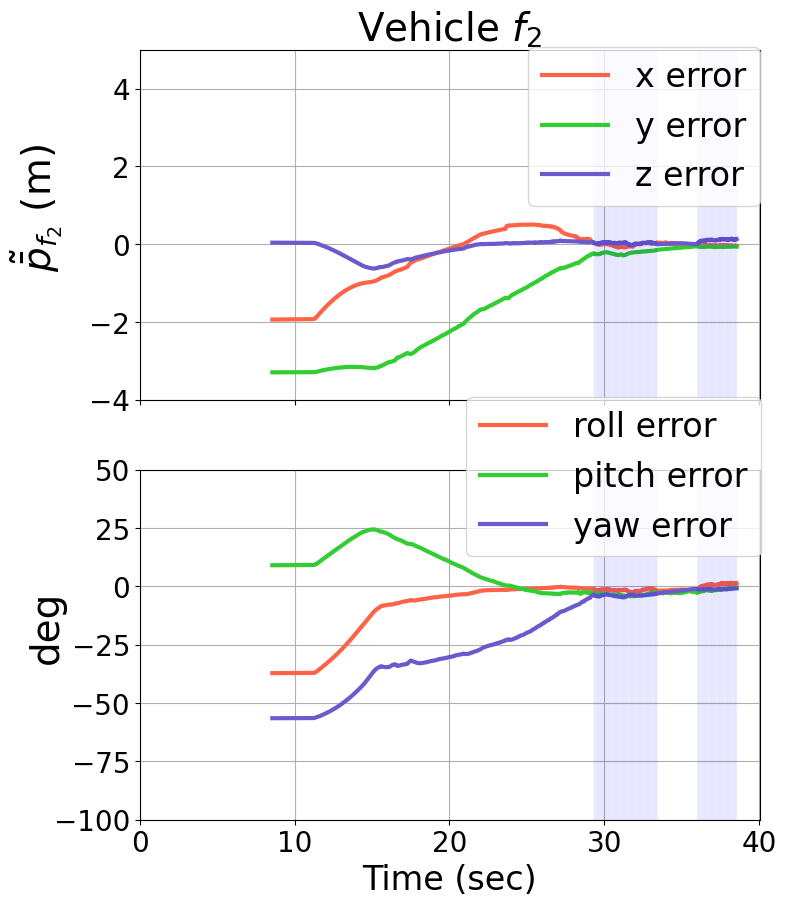}
         \label{fig:body_overtake_f2}
         }
     \end{minipage}
     \begin{minipage}[t]{1\columnwidth}
     \subfigure[]{
        \includegraphics[width=0.45\linewidth]{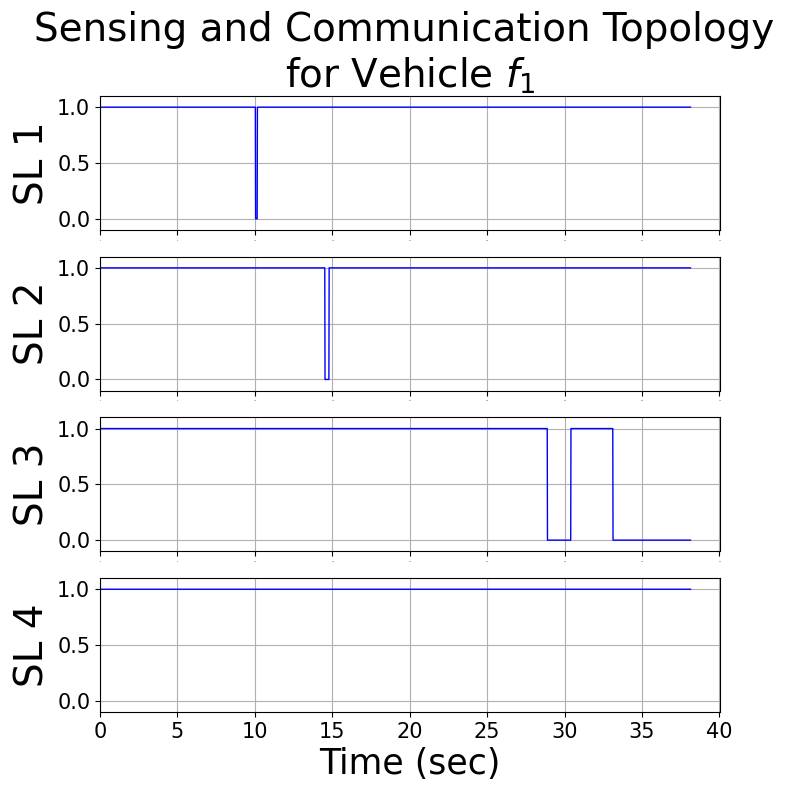}
        \label{fig:body_overtake_landmark_detection_f1}
     }
     \subfigure[]{
        \includegraphics[width=0.45\linewidth]{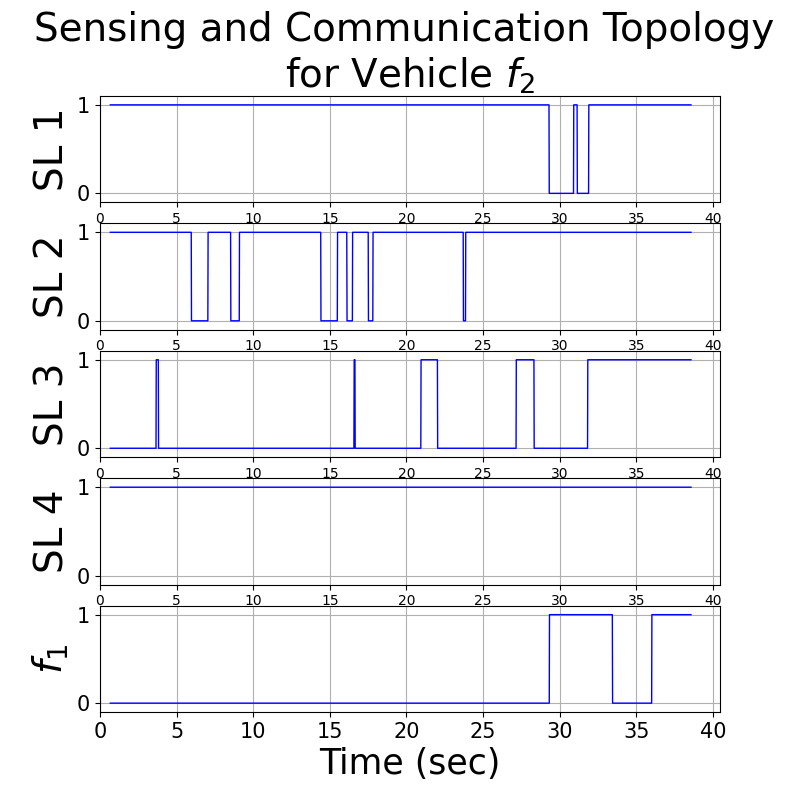}
        \label{fig:body_overtake_landmark_detection_f2}
     }
     \end{minipage}
     \vspace{-1.5em}
    \caption{(a) The projected 2D trajectories of vehicles $f_1$ and $f_2$'s position estimations and ground truth in the overtaking scenario, with triangles and crosses indicating the initial and final positions, respectively, dotted and solid lines indicating the ground truth and estimation, respectively, SL refers to static landmark.
    (b) Evolution of the estimation errors of Vehicle $f_1$. (c) Evolution of the estimation errors of vehicle $f_2$.
    (d) Sensing and communication topology of vehicle $f_1$. 
    (e) Sensing and communication topology of vehicle $f_2$. 
    }
    \label{fig:body_overtake}
     \vspace{-1.5em}
\end{figure}

\subsubsection{\textbf{Overtaking Scenario}} as illustrated in Fig.~\ref{fig:overtaking_intersection_setup}, vehicle $f_1$ overtakes vehicle $f_2$ as they approach and enter the intersection. This behavior occurs when vehicles intend to overtake slower vehicles or to change lanes for a turn. In this scenario, we consider a $6$-agent system: there are four static landmarks, and two vehicle agents, $f_1$ and $f_2$, that need to estimate their poses. The gains used for vehicle agents $f_1$ and $f_2$ are the same as in the crossing path scenario.

Fig. \ref{fig:body_overtake_2d} shows the projected 2D trajectories (estimated and ground truth) of the vehicles.
Figs. \ref{fig:body_overtake_f1}, and \ref{fig:body_overtake_f2} show the convergence of the estimation errors of vehicles $f_1$ and $f_2$ throughout the experiments, respectively. The blue-shaded areas in Fig. \ref{fig:body_overtake_f2} represent when vehicle $f_2$ detects vehicle $f_1$. 
We observe that the steady-state orientation estimation error of vehicle $f_1$ is slightly bigger than that of vehicle $f_2$, as shown in Fig. \ref{fig:body_overtake_f1}.  This is due to the lateral tilt of vehicle $f_1$ caused by the uneven tire pressure. 
Figs. \ref{fig:body_overtake_landmark_detection_f1} and \ref{fig:body_overtake_landmark_detection_f2} show the interaction topologies of vehicles $f_1$ and $f_2$, respectively. They are both switching graph topologies due to the limited camera's field of view and occlusion. Occlusions occurred for vehicle $f_2$ from 29s to 34s, see Fig. \ref{fig:body_overtake_landmark_detection_f2}, when vehicle $f_1$ overtook vehicle $f_2$, which blocked up to 2 static landmarks. 
We conclude that the proposed decentralized observer is effective under a leader-follower structure. The experimental video of this scenario can be found at: 
\href{https://bit.ly/overtaking-scenario}{https://bit.ly/overtaking-scenario}.

\begin{figure}
    \centering
    \vspace{0.5em}
    \includegraphics[width=0.8\linewidth]{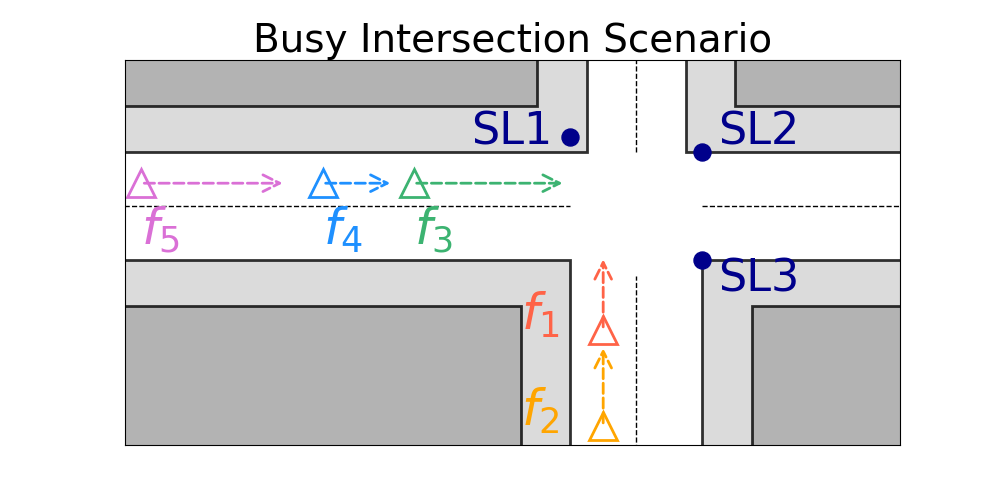}
    \vspace{-0.5em}
    \caption{Simulation Scenario of a busy intersection, with SL referring to static landmark, and $f_1$ - $f_5$ denote the vehicles estimating their poses, with triangles denoting the initial positions and arrows showing the direction of motion.}
    \label{fig:simulation-5-vehicles-topology}
    \vspace{-1.75em}
\end{figure}

\subsubsection{\textbf{Simulation for busy intersection scenario}}
Due to space limitations in the experimental setup, the experiments were conducted using only two vehicle agents. Nevertheless, to demonstrate the scalability of the proposed cooperative localization approach, we present a simulation illustrating its potential applicability to scenarios involving a larger number of vehicles at busy intersections.  We consider an $8$-agent system as shown in Fig. \ref{fig:simulation-5-vehicles-topology}: three static landmark agents scattered around the intersection at $\bar p_1 = (-4, 5, 3)$, $\bar p_2 = (4, 4, 5)$ and $\bar p_3 = (4, -3, 4)$, two vehicle agents $f_1$, $f_2$ traveling northbound  and  three other vehicle agents $f_3$, $f_4$, and $f_5$ move eastbound.


The trajectories of the vehicles are chosen as $\bar p_{f_1}(t) = (-2,-16 + 0.6t,2.5)^\mathsf{T}$, $\bar p_{f_2}(t) = (-2,-19+0.5t,2)^\mathsf{T}$, $\bar p_{f_3}(t) = (-17+0.6t,2,3)^\mathsf{T}$, $\bar p_{f_4}(t) = (-19+0.45t,2,3.5)^\mathsf{T}$, and $\bar p_{f_5}(t) = (-30+0.6t,2,3)^\mathsf{T}$, with $R_{f_i}(t)=I_3$.
The initial estimations are chosen as $\hat {\bar p}_{f_1}(0) = (0, -5, 5)^\mathsf{T}$, $\hat {\bar p}_{f_2}(0) = (5, -14, 6)^\mathsf{T}$, $\hat {\bar p}_{f_3}(0) = (-8, 3, 5)^\mathsf{T}$, $\hat {\bar p}_{f_4}(0) = (-14, 6, 6)^\mathsf{T}$, $\hat {\bar p}_{f_5}(0) = (-24, 7, 6)^\mathsf{T}$, and $\Lambda_{f_i}(0) = (\frac{\sqrt{2}}{2}, 0, 0,\frac{\sqrt{2}}{2})^\mathsf{T}$. The interaction topologies are designed as: $\mathcal N_{f_1}=\{1,2,3\}$, $\mathcal N_{f_2}=\{2,3, f_1\}$,  $\mathcal N_{f_3}=\{2,3, f_2\}$, $\mathcal N_{f_4}=\{3,f_2,f_3\}$, $\mathcal N_{f_5}=\{f_1,f_2,f_4\}$. The gains are chosen as $k_i = 1$, $q_{ij}=10 \quad \forall j \in \mathcal{N}_i$, $V_i=\diag\{0.1I_3, I_3\}$, and $P_i(0) =\diag \{I_3, 100I_3\}$. 

To validate the robustness of our method, we consider noisy measurements for translational 
and angular velocities, as well as inter-agent bearings, as in \cite{hamel2017riccati}.  Zero-mean Gaussian noises are added with a standard deviation of 0.1 for translational
velocity measurements $v_i(t) + b_{v}(t)$ and a standard deviation of 0.01 for angular velocity measurements $\omega_i(t) + b_{\omega}(t), \forall i\in \mathcal V_f$. For bearing measurements, we use the noisy bearing defined as 
\begin{equation}
    \begin{aligned}
    g_{ij}^n =sign(g_{ij,3}) \frac{{p_{ij}'}}{||p_{ij}'||} 
    \end{aligned}
\end{equation}
with $$p_{ij}'=(g_{ij,1}/g_{ij,3}+n_1, g_{ij,2}/g_{ij,3}+n_2,1)^\top$$
and $n_1$ and $n_2$ zero-mean, uniformly distributed noises with a maximum deviation of 0.005.

\begin{figure}
    \centering
    \vspace{0.5em}
    \includegraphics[width=0.98\linewidth]{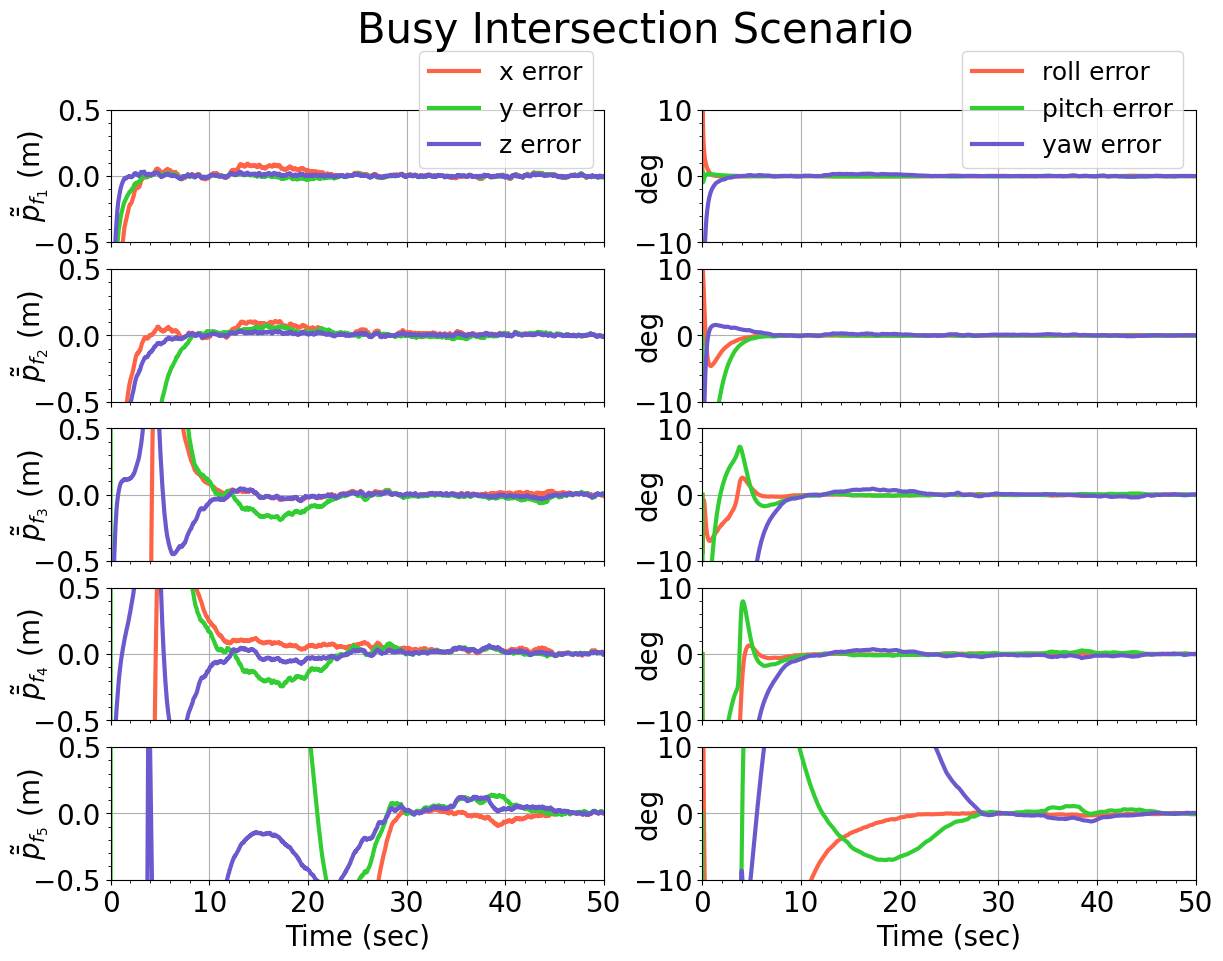}
    \vspace{-1.0em}
    \caption{Evolution of error states (position in common inertial frame $\mathcal{F}$ and orientation) of vehicles $f_1$ - $f_5$ (from top to bottom).} 
    \label{fig:simulation-5-vehicles-errors}
\end{figure}


\begin{figure}
    \centering
    \includegraphics[width=0.8\linewidth]{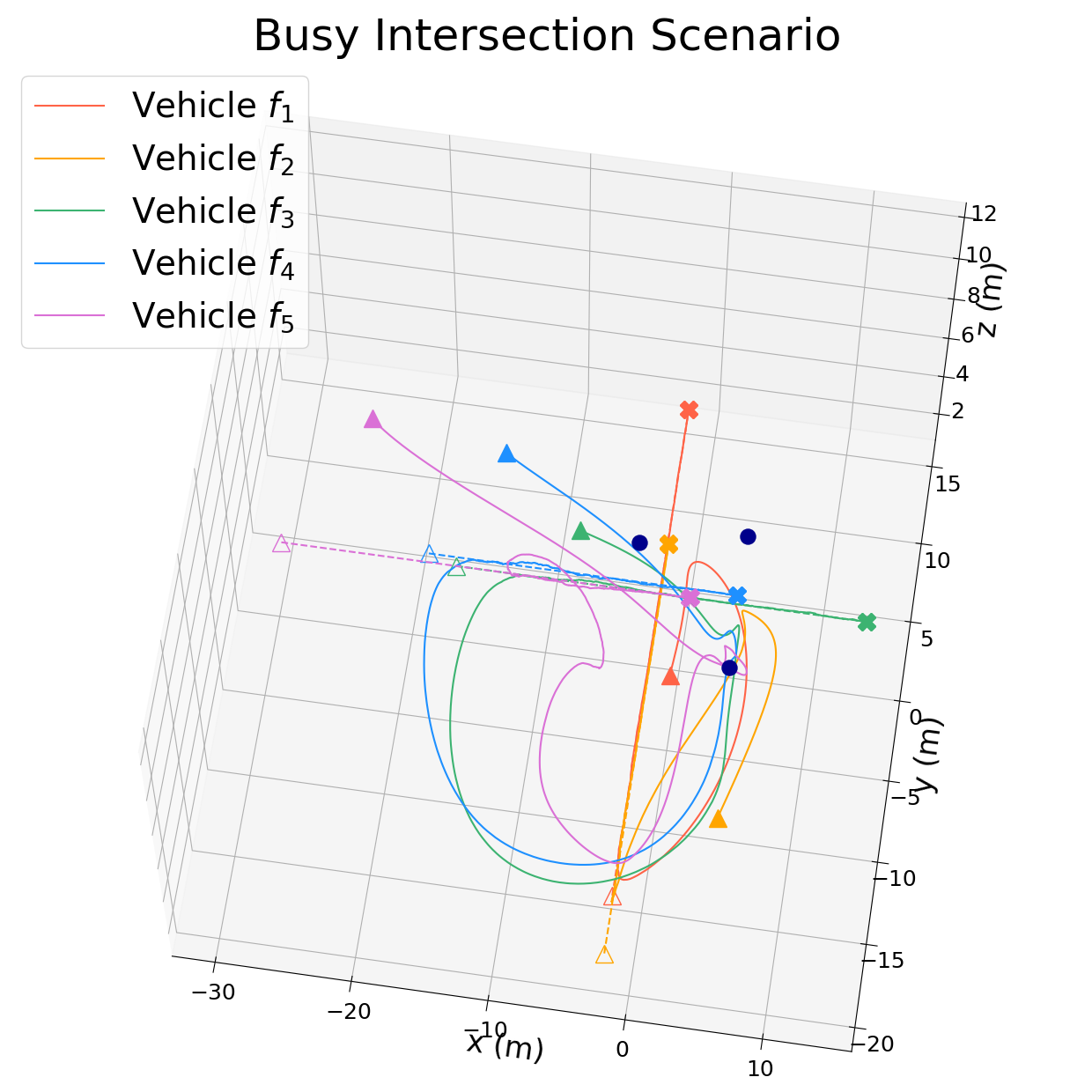}
    \vspace{-1em}
    \caption{The 3D trajectories of vehicles $f_1$ - $f_5$ with triangles and crosses indicating the initial and final poses, respectively, dotted and solid lines indicating the ground truth and estimation, respectively, and darkblue dots indicating the static landmark positions.}
    \label{fig:simulation-5-vehicles-3d}
     \vspace{-0.5em}
\end{figure}

Fig. \ref{fig:simulation-5-vehicles-errors} shows the convergence of the estimation errors to zero for vehicles $f_1$ - $f_5$, and Fig. \ref{fig:simulation-5-vehicles-3d} shows the 3D trajectories of the five vehicles during the simulation. The simulation results validate the scalability and effectiveness of the proposed decentralized observer under measurement noises. 

\section{Conclusion}\label{sec:conc}

In this paper, we propose a decentralized observer for a group of connected and automated vehicles using relative bearing measurements from onboard cameras and translational and angular velocity measurements. The proposed method addresses the issue of cameras' occlusion caused by neighboring vehicles in complex urban environments, such as intersections, by leveraging V2V and V2I communication. This enables the ego-vehicle to utilize occluding vehicles as landmarks. 
We prove the exponential stability of the estimation error's origin, validate and showcase the robustness and performance of the proposed cooperative localization approach on both real
hardware and in larger-scale simulations.
For future work, we will formally analyze the minimal observability condition mentioned in the Remark \ref{remark} and evaluate the practical implementation of the presented method on experimental 5G / 6G testbeds, such as~\cite{arfvidsson2024small}.

\section{Appendix}\label{sec:appex}

\subsection{Rotation matrix}\label{appex:rotation_body}
The corresponding Rodrigues formula relating $\Lambda_i$ to $\tilde R_i$ is
$$\tilde{R}_i=I_3+2S(\lambda_i)(\mathring{\lambda}_i I_3+S(\lambda_i))$$
Define $\tilde{\omega}_i:=\omega_i-\tilde{R}_i^\mathsf{T}\hat \omega_i$, we can deduce the following kinematic relations
$$\dot{\tilde{R}}_i=\tilde{R}_i S(\tilde{\omega}_i);\ \left\{ \begin{aligned}
\dot{\mathring{\lambda}}_i&=-0.5\tilde{\omega}_i^\mathsf{T}\lambda_i \\
\dot{\lambda}_i&=0.5\tilde{\omega}_i\mathring{\lambda}_i+0.5S({\tilde{\omega}}_i)\lambda_i.
\end{aligned}
\right.$$
Since $\mathring{\lambda}_i=1-|\lambda_i|^2$ and $\omega_i$ is bounded, we have \eqref{eq:Inertialdotlambdabody}.

\subsection{Representation of measurement $y_{ij}$} \label{appex:y-body}
Using the fact
\begin{equation*}\scalebox{0.95}{$
    \begin{aligned}
        |p_i - R_i^\mathsf{T}\bar p_j|g_{ij} &= p_i - R_i^\mathsf{T}\bar p_j \\
        &= p_i - \tilde{R}_i^\mathsf{T}\hat{R}_i^\mathsf{T}\bar p_j \\
        &= p_i - \hat{R}^\mathsf{T}_i\bar p_j - (\tilde{R}_i^\mathsf{T} - I_3)\hat{R}_i^\mathsf{T} \bar p_j \\
        &= p_i - \hat{R}^\mathsf{T}_i\bar p_j + S(2\lambda_i)\hat{R}_i^\mathsf{T}\bar p_j + O(|\bar p_j||\lambda_i|^2) \\
        &= p_i - \hat{R}^\mathsf{T}_i\bar p_j - S(\hat{R}_i^\mathsf{T}\bar p_j)(2\lambda_i) + O(|\bar p_j||\lambda_i|^2) \\
        &= p_i - \hat{R}^\mathsf{T}_i\bar p_j - S(\hat{R}_i^\mathsf{T} \hat R_j \hat p_j)(2\lambda_i) \\ 
        & \quad + O(|\bar p_j||\lambda_i|^2) + O(|\bar p_j-\hat R_j \hat p_j||\lambda_i|)
    \end{aligned}$}
\end{equation*}
together with $\Pi_{g_{ij}}g_{ij}=0$ and $|p_i-R_i^\top\bar p_j|$ is bounded, one has
\begin{equation*}
    \begin{aligned}
    \Pi_{g_{ij}} \hat R_i^\mathsf{T} \bar p_j &= \Pi_{g_{ij}}p_i - \Pi_{g_{ij}}S(\hat R_i^\mathsf{T} \hat R_j \hat p_j)(2\lambda_i) \\
    & \quad + O(|\bar p_j||\lambda_i|^2) + O(|\bar p_j-\hat R_j \hat p_j||\lambda_i|)
    \end{aligned}
\end{equation*}

Hence, we can represent $y_{ij} = \Pi_{g_{ij}} \hat R_i^\mathsf{T} \hat R_j\hat p_j$ as

\begin{equation*}\scalebox{0.95}{$
    \begin{aligned}
        y_{ij} &= \Pi_{g_{ij}} \hat R_i^\mathsf{T} \bar p_j - \Pi_{g_{ij}} \hat R_i^\mathsf{T} (\bar p_j-\hat R_j \hat p_j)\\
        &= \Pi_{g_{ij}}p_i - \Pi_{g_{ij}}S(\hat R_i^\mathsf{T} \hat R_j \hat p_j)(2\lambda_i) - \Pi_{g_{ij}}\hat R_i^\mathsf{T} (\bar p_j-\hat R_j \hat p_j) \\
    & \quad + O(|\bar p_j||\lambda_i|^2) + O(|\bar p_j-\hat R_j \hat p_j||\lambda_i|)
    \end{aligned}$}
\end{equation*}

\subsection{Proof of Theorem \ref{theorem:mobilevelocity}} \label{appex:proof}
Define $x_i = \begin{bmatrix}
    2\lambda_i^\mathsf{T} & p_i^\mathsf{T}
\end{bmatrix}^\mathsf{T}$ 
and recall \eqref{eq:dynamics-body}, we have 
\begin{equation}
    \dot {x}_i = A_i x_i + \begin{bmatrix}
        \tilde \omega_i \\
        v_i
    \end{bmatrix} + \begin{bmatrix}
        O(|\lambda_i||\tilde \omega_i|) + O(|\lambda_i|^2) \\ 
        0
    \end{bmatrix}
\end{equation}
where $A_i = \begin{bmatrix}
        -S(\omega_i) & 0_3 \\
        0_3 & -S(\omega_i)
    \end{bmatrix}$. Define the estimate of $x_i$ as $\hat x_i = \begin{bmatrix}
    0_3^\mathsf{T} & \hat{p}_i^\mathsf{T}
\end{bmatrix}^\mathsf{T}$ and recall \eqref{eq:observer_body}, we have
\begin{equation}
    \dot{\hat {x}}_i = A_i \hat{x}_i + \begin{bmatrix}
         \tilde \omega_i \\
         v_i
    \end{bmatrix} + K_i(y_i - C_i\hat x_i)
\end{equation}

Denote $\tilde x_i = x_i - \hat x_i$ and we have
\begin{equation}\scalebox{0.9}{$
    \dot {\tilde {x}}_i = A_i\tilde {x}_i - K_iC_i\tilde{x}_i + K_iD_i(\tilde x_j^{'}) + O(|\tilde{x}_i|^2) + O(|\tilde {x}_i||\tilde {x}_j^{'}|), \forall j \in \mathcal{N}_i$}
\end{equation}
where $\tilde x_j^{'}:=[0_3^\top \ (\bar p_j-\hat R_j \hat p_j)^\top]^\top$.

We will first prove that the theorem is true for the 1st and the 2nd vehicle agents, and then we will show that it is true for all $i \in \mathcal{V}_f$ using mathematical induction. 

Proof for the 1st vehicle agent:

Denote $f_1 = n_l + 1$ as the vertex index of the 1st vehicle agent. Since $\mathcal{N}_{f_i} \subseteq \mathcal{V}_l$ and $\tilde x_j = 0$, $\forall j \in \mathcal{V}_l$, we have
\begin{equation}
    \dot {\tilde{x}}_{f_1} = (A_{f_1}  -K_{f_1}C_{f_1})\tilde{x}_{f_1} + O(|\tilde{x}_{f_1}|^2).
\end{equation}
Since $(A_{f_1}, C^*_{f_1})$ is uniformly observable and $Q_{f_1}$ is larger than some positive matrix, we conclude that $\tilde x_{f_1} = 0$ is ES by a direct application of \cite[Thereom 3.1]{hamel2017riccati}.

Proof for the 2nd vehicle agent:

Denote $f_2 = n_l +2$ as the vertex index of the 2nd vehicle agent. If $f_1 \notin \mathcal{N}_{f_2}$, the proof is identical to the 1st vehicle agent. If $f_1 \in \mathcal{N}_{f_2}$, we have a system
\begin{equation}\label{eq:f2_eq}
    \begin{aligned}
    \dot{\tilde{x}}_{f_2} &= (A_{f_2} - K_{f_2}C_{f_2})\tilde {x}_{f_2} + K_{f_2}D_{f_2}(\tilde x_{f_1}) \\
    & \qquad + O(|\tilde{x}_{f_2}|^2) + O(|\tilde {x}_{f_2}||\tilde {x}_{f_1}|)
    \end{aligned}
\end{equation}
which can be considered as a cascaded system that has $\tilde {x}^{'}_{f_1}$ as a perturbation to the unforced system
\begin{equation}\label{eq:f2_eq_unforced}
    \dot{\tilde{x}}_{f_2} = (A_{f_2} - K_{f_2}C_{f_2})\tilde {x}_{f_2} + O(|\tilde{x}_{f_2}|^2)
\end{equation}
and by a direct application of \cite[Thereom 3.1]{hamel2017riccati}, we can conclude that the equilibrium $\tilde {x}_{f_2}=0$ of the unforced system \eqref{eq:f2_eq_unforced} is ES. Recall \eqref{eq:D_body}, since $\Pi_{g_{f_2f_1}}\hat R^\mathsf{T}_{f_1}$ is bounded by definition, 
and $\tilde {x}_{f_1} = 0$ is ES, so is $\tilde {x}_{f_1}^{'}$, this implies that the equilibrium point $\tilde {x}_{f_2} = 0$ is ES for the system \eqref{eq:f2_eq}. 

Proof for the $n_f$th ($n_f\ge 3$) vehicle agents:

We have proved the theorem is true for the 1st and 2nd vehicle agents. Now we assume that it is true for the $(k-1)$th vehicle agent with $3 \leq k-1 \leq n_f-1$, and we will prove that it is true for the $k$th vehicle agent. Denote that $f_k = n_l+k$ as the vertex index of the $k$th vehicle agent. The proof follows a similar argument to that of the 2nd vehicle agent. If $\mathcal{V}_f \cup \mathcal{N}_{f_k} = \emptyset$, the proof is identical to the 1st vehicle agent. If $\mathcal{V}_f \cup \mathcal{N}_{f_k} \neq \emptyset$, then we have a system
\begin{equation}\label{eq:fk_eq}
    \begin{aligned}
    \dot {\tilde {x}}_{f_k} &= (A_{f_k} - K_{f_k}C_{f_k})\tilde {x}_{f_k} + K_{f_k}D_{f_k}(\tilde x_{f_j}^{'}) \\
    & \qquad + O(|\tilde{x}_{f_k}|^2) + O(|\tilde {x}_{f_k}||\tilde {x}_{f_j}^{'}|)
    \end{aligned}
\end{equation}
which can be considered as a cascaded system that has $\tilde {x}_{f_j}^{'}$, $\forall j \in \mathcal{V}_{f_k}$, and $f_j \geq f_1$ as perturbations to the unforced system 
\begin{equation}\label{eq:fk_eq_unforced}
    \dot {\tilde {x}}_{f_k} = (A_{f_k} - K_{f_k}C_{f_k})\tilde {x}_{f_k} + O(|\tilde{x}_{f_k}|^2)
\end{equation}
and by a direct application of \cite[Thereom 3.1]{hamel2017riccati}, we can conclude that the equilibrium $\tilde {x}_{f_k}=0$ of the unforced system \eqref{eq:fk_eq_unforced} is ES. Recall \eqref{eq:D_body}, since $\Pi_{g_{f_kf_j}}\hat R^\mathsf{T}_{f_k}$ is bounded by definition, and $\tilde {x}_{f_j} = 0, \forall f_j \in \mathcal{N}_{f_k}$ is ES, so is $\tilde x_{f_j}^{'}$, this implies that the equilibrium point $\tilde {x}_{f_k} = 0$ is ES for the system \eqref{eq:fk_eq}.

\balance






\end{document}